\documentstyle[aps,eqsecnum,epsf]{revtex}

\begin{document}

\draft

\title{Channel Flow of Smectic Films}
\author{Shilpa Jain and David R. Nelson}
\address{Lyman Laboratory of Physics, Harvard University, Cambridge, Massachusetts 02138}
\maketitle

\begin{abstract}

The hydrodynamics of smectic films at an air-water interface is discussed, with particular focus on the viscous response of the film under flow normal to the layers. The corrections to the response functions of the smectic phase, arising from the coupling between the flow and the smectic order parameter, are calculated. The results for the effective viscosity are illustrated by analysing smectic film flow in a channel geometry. Two limiting cases of the flow, namely, motion dominated by dislocation-induced shear-softening and dislocation-free motion dominated by the permeation mode of mass transfer, are studied. The effect of drag from a finite depth liquid subphase is considered. The results are compared to those for hexatic and liquid films.

\end{abstract}

\section{Introduction}

The rheological properties of liquid crystalline systems continue to be of considerable interest because of their rich and complex behaviour. Thin films of anisotropic molecules, such as Langmuir monolayers at an air-water interface, are relevant to many industrial applications, and as such, have been subjected to detailed experimental studies. In particular, the viscous response of liquid crystalline films is often found to be non-Newtonian. Among the dominant causes is coupling of the flow to molecular alignment. In smectic films (crystalline in one dimension, but liquid-like in the other), the presence of unbound dislocations becomes a major factor affecting viscous response. When the film is riding over another phase, usually water, viscous drag from this subphase, if large, can modify the flow profile of the film quite significantly.

The coupling between molecular alignment and flow has been seen in some cases in experiments by Mingotaud \textit{et al.}~\cite{orient}, Maruyama \textit{et al.}~\cite{Fuller}, and Kurnaz \& Schwartz~\cite{Schwartz}. The experiments involve Langmuir monolayers of rod-shaped molecules that are usually tilted with respect to the surface normal, forming a hexatic phase with anisotropic in-plane bond orientations. Typically the film consists of domains of a liquid crystalline phase co-existing with another liquid crystalline phase or with the liquid expanded phase (no orientational order). The domains can be distinguished through Brewster Angle Microscopy which is sensitive to molecular orientation, making it possible to follow the shape and movement of the domains along the flow. There is evidence of nonlinear shear response~\cite{Schwartz,elastic} emerging from such studies, as well as of the molecular orientation being influenced by flow~\cite{orient,Fuller}.

Shear-thinning has often been observed in experiments involving Langmuir monolayers~\cite{Schwartz,elastic}. A possible explanation of this phenomenon is provided by Bruinsma \textit{et al.}~\cite{shear-thinning} in terms of shear-induced defect proliferation. Dislocation defects in a solid, if unbound, can relax an applied strain by moving in response to the resulting stress. Since the force on such a defect depends on its ``charge'', oppositely charged defects tend to separate under an external stress. Tightly bound pairs cannot contribute to the steady state viscous response, although they can modify the response at non-zero frequencies and wave-vectors~\cite{scresponse}. However, at finite temperatures, bound pairs can dissociate under this separating influence which effectively tilts the potential well confining the pair, allowing them to be free beyond a potential barrier. Being thermally activated, the free dislocation density, and hence the viscous rsponse, would be very sensitive to temperature. Experiments conducted by Schwartz \textit{et al.}~\cite{private} on anisotropic hexatic and crystalline phases of Langmuir monolayers do indeed see a strong temperature dependence of the critical shear rate for onset of non-Newtonian behaviour.

In this paper we study a simpler problem, the linear hydrodynamics of two-dimensional smectic films in a channel flow geometry. Dislocations still play an important role, and it is easier to analyze their effect on the smectic order embodied in a \textit{single} set of Bragg planes. Although we do not study this here, channel flow of two-dimensional smectics would also be a promising context in which to explore a tractable model of shear thinning.

In the absence of external strains, free dislocations or disclinations don't occur in the most ordered two-dimensional phases; they are instead bound in pairs of opposite charges by a logarithmic potential. However, in two-dimensional layered materials such as smectics or cholesterics, there is exponential decay of translational order in both the layering direction, and the liquid-like direction along the layers~\cite{Toner}. As a result, isolated dislocations have a finite energy and exist in a finite concentration at any finite temperature. In these materials, shear response is primarily due to the free dislocations; the viscosity diverges inversely as the dislocation density when it becomes small at low temperatures. This divergence is cut off at short length scales by the permeation mode of mass transfer in smectics, where a layer distortion induces molecules to jump from layer to layer without affecting the layering structure, allowing the distortion to relax over a finite distance. An analogy can be made to the screening effect of supercurrents in a superconductor that allow the magnetic field to penetrate a finite distance in from the surface, although the field is expelled from the bulk. The ``permeation current'' plays a similar role with respect to shear in a smectic without dislocations.

In three dimensions, this analogy has been carried further to predict divergence of the response functions of the smectic near the second-order smectic-to-nematic transition. The coupling of the nematic order parameter to fluctuations in the magnitude of the smectic order parameter causes, among other quantities, the permeation constant of the smectic and the viscosity denoted $\eta_1$ in the literature, to diverge~\cite{critical}. However, in two dimensions, a dislocation-driven thermodynamic transition to the nematic state occurs at zero temperature~\cite{Toner}, dislocations being the analog of Abrikosov flux vortices in two-dimensional superconductors. At a finite temperature, the nematic melts into an isotropic liquid via a disclination unbinding transition. Below this temperature, local smectic order is disrupted by singularities in the phase of the order parameter (\textit{i.e.}, dislocations). However, the local smectic order parameter has a finite magnitude, and fluctuations in the magnitude are irrelevant in the renormalization sense~\cite{scaling}. Thus renormalization of the elastic coefficients and response functions as in the 3-d case does not occur. The phase of the local smectic order parameter is related to the layer displacement, and dislocations, which lead to branch cuts in this phase, cost a finite energy as for magnetic vortices in the 3-d superconductor.

Coupling of the film flow to a subphase (a fluid body supporting the film on its surface) can also significantly alter its flow profile. Such experiments have been conducted by Schwartz \textit{et al.}~\cite{sub,Schwartz} using Langmuir monolayers on water. When the subphase drag dominates the flow, the flow profile becomes semi-elliptical. Stone~\cite{Stone} has performed calculations which confirm this profile and also yield the profiles interpolating between the elliptical and the parabolic as the viscosity of the film relative to that of the subphase is increased. The depth of the subphase was also a parameter in the calculations, since decreasing the depth results in increased drag. As we show below, Stone's results can also be applied to hexatic films.

The hydrodynamics of partially ordered hexatic films has been studied in detail by Zippelius \textit{et al.}~\cite{hydro}. These authors find a correction to the effective viscosity under flow conditions where the hexatic bond orientation is pinned at the boundaries, as compared to the case where the bond orientation is free to rotate. The correction comes from coupling of the flow to the bond orientation order parameter under the constraint imposed by the boundaries.

A similar coupling can be enforced by imposing a pressure gradient on the flow. In the experiments conducted by Kurnaz \& Schwartz~\cite{Schwartz} on hexatic film flow, the domain structure of the hexatic mesophase can impose constraints on the bond orientation at domain boundaries, thus increasing the viscosity from its bare value. Annealing of the domains would then lead to a reduction in the effective viscosity. The experimental signature of this effect would be a time-dependent viscous response. Some transients have indeed been observed in these experiments, although other factors may be involved, such as domain boundary elasticity~\cite{private} and shear thinning.

To illustrate how to incorporate effects of a subphase into the hydrodynamics of a partially ordered film, we adapt the analysis of Ref.~\cite{Stone} to films with hexatic order. In the presence of a subphase, the hydrodynamic equations of motion for a hexatic get modified by adding a subphase drag term to the viscous force: denoting by $x$ the co-ordinate across the channel, $z$ along the channel, and $y$ along the channel depth (see Fig.~\ref{channel-sub}), we have (with stress matrix $\sigma_{ij}$),
\begin{mathletters}
\label{hex}
\begin{eqnarray}
\label{g}
\frac{\partial{g_z}}{\partial{t}} &=& \partial_z \sigma_{zz} + \partial_x \sigma_{zx} + \partial_y \sigma_{zy} \nonumber\\
 &=& \pi' + \partial_x \left( \nu \partial_x g_z - \frac{K_A}{2} \partial_x^2 \theta \right) - \nu_b \, \partial_y g_z|_{\mathit film\ surface} \\
\label{th}
\frac{\partial{\theta}}{\partial{t}} &=& \frac{\partial_x g_z}{2 \rho} + \Gamma_6 K_A \partial_x^2 \theta
\end{eqnarray}
\end{mathletters}
where~\cite{hydro} $\mathbf{g}$ is the momentum density, $\theta$ is the hexatic bond orientation order parameter, $K_A$ the bond orientation stiffness, $\Gamma_6$ the corresponding kinetic coefficient, $\nu$ the dynamic shear viscosity ($\nu = \eta/\rho$), $\nu_b$ the viscosity of the bulk subphase, and $\pi'$ the surface pressure gradient driving the film down the channel.

Assuming $\partial_x \theta$ is time-independent in the steady state, we have $\partial_t \partial_x \theta = 0$. Therefore $\partial_t \theta$ must be constant across the channel. Since $g_z$ is an even function of $x$ in this flow situation, $\partial_t \theta$ is odd, and hence, must be zero. Eq.~(\ref{th}) then gives us the coupling between the flow and the bond orientation:
\begin{equation}
\partial_x^2 \theta = - \frac{1}{2 \rho \Gamma_6 K_A} \partial_x g_z .
\end{equation}
Upon substituting this result into Eq.~(\ref{g}), we find:
\begin{equation}
\frac{\partial{g_z}}{\partial{t}} = \pi' + \partial_x \left( \left( \nu + \frac{1}{4 \rho \Gamma_6} \right) \partial_x g_z \right) - \nu_b \, \partial_y g_z|_{\mathit film\ surface} .
\end{equation}
Thus $g_z$ obeys an equation of motion identical to that for an isotropic film with a subphase, but with the modified viscosity $\eta + 1/4 \Gamma_6$, and we can take over the results of Ref.~\cite{Stone}. As we shall see (Sec.~\ref{subphase}), this simplification does not apply to smectic flow.

In the next section, we briefly review the equilibrium properties of two-dimensional smectic films~\cite{Toner}. Section \ref{normal} discusses the hydrodynamics of two-dimensional smectics, and the implications of the coupling between the smectic order parameter and the flow for the response functions of both quantities. Section \ref{channel} looks at flow of a smectic film in a channel flow geometry, and examines the behaviour in different regimes of the channel width. In Section \ref{subphase}, we consider the effect of subphase drag on the smectic film flow as compared to previous results for an isotropic film. The results of both sections \ref{channel} and \ref{subphase} are consistent with the results of section \ref{normal} for the effective viscosity. The last section summarizes the results of this paper.

\section{Review of Smectic Films}
\label{review}

Smectics are characterized by a crystal-like periodic modulation of the density along one direction, say, the $z$-direction, and liquid-like correlations perpendicular to it. In two dimensions, we take this to be the $x$-direction. The preferred orientation of the ``layers'' is also the average direction along which the directors $\hat{\mathbf n}$ of the nematic molecules are oriented. Although it represents a spontaneously broken rotational symmetry, the layer orientation can be forced by boundary conditions on the molecules, or even by flow. Smectic order is characterized by a wavevector ${\mathbf q}_0 = \hat{\mathbf z} 2 \pi /d$, where $d$ is the layer spacing, usually slightly larger than the molecular length.

The smectic density wave can be represented as~\cite{prost}
\begin{equation}
\rho({\mathbf r}) = \rho_0 \left( 1 + \psi({\mathbf r}) e^{i {\mathbf q}_0 \cdot {\mathbf r}} \right) .
\end{equation}
Here, $\psi({\mathbf r})$ is the complex smectic order parameter: its amplitude represents the strength of the smectic ordering, whereas the phase $\phi({\mathbf r}) = q_0 u({\mathbf r})$ describes the phonons associated with broken translational symmetry along the layering direction. Phonons in two dimensions are very effective in destroying the one-dimensional translational order: The correlation $\langle \psi({\mathbf r}) \psi^*({\mathbf 0}) \rangle$ decays as the exponential of a power of the displacement. Since the square of the wavevector appears in the exponent~\cite{Toner}, higher harmonics of $q_0$ in the density modulation are ignored.

The Landau-Ginzburg free energy takes the form~\cite{gennes}
\begin{equation}
{\mathcal F} = \int\!\! d^{2}r \left[ \frac{a}{2} |\psi|^2 + \frac{u}{4} |\psi|^4 + \frac{c_\parallel}{2} |\partial_z \psi|^2 + \frac{c_\perp}{2} |(\partial_x - i q_0 \delta n) \psi|^2 + \frac{K_1}{2} (\partial_x \delta n)^2 + \frac{K_3}{2} (\partial_z \delta n)^2 \right] .
\end{equation}
Here, $K_1$ and $K_3$ are splay and bend elastic constants. The twist elastic constant $K_2$ is absent in two dimensions. The coupling between $\partial_x \psi$ and $\delta n \equiv (\hat{\mathbf n}-\hat{\mathbf z}) \cdot \hat{\mathbf x}$ is required to satisfy the rotational invariance of ${\mathcal F}$. Terms of order higher than quadratic in the order parameter and its gradients have been neglected. Well below the mean-field smectic-nematic transition temperature, fluctuations in the amplitude of $\psi$ can also be ignored, and in the absence of singularities in $\hat{\mathbf n}$ (disclinations), $\delta n$ can be integrated out.The remaining long wavelength fluctuations can be expressed completely in terms of the layer displacement $u({\mathbf r})$ as~\cite{prost,Toner}
\begin{equation}
{\mathcal F} = \int\!\! d^{2}r \frac{1}{2} B \left[ (\partial_z u)^2 + \lambda^2 (\partial_x^2 u)^2 \right] ,
\end{equation}
with $B = \psi_0^2 q_0^2 c_\parallel$, and $\lambda^2 = K_1/B$. Note that uniform gradients of $u$ along the layer direction ($\partial_x u$) don't cost any energy, because they represent tilting of the layering direction. This important difference compared to two-dimensional solids, hexatics, etc. implies that the lowest energy defects in the system, dislocations, have a \textit{finite} energy $E_D$ and are not constrained to be bound in pairs at low temperatures~\cite{Toner}.

Whereas a smectic with thermally excited phonons would behave like a nematic with only a splay degree of freedom, the presence of dislocations allows for bend in the average layer orientation over scales larger than the typical size $\xi_D$ of a correlated ``smectic blob''~\cite{Toner}, given by $\xi_D^2 \equiv n_D^{-1} \approx a_D^2 e^{E_D/k_B T}$ ($a_D$ is a dislocation core diameter, $a_D^2 \sim d \sqrt{\lambda d}$). Therefore the long-wavelength behaviour of the smectic is that of a nematic with free energy
\begin{equation}
\label{fn}
{\mathcal F} = \int\!\! d^{2}r \frac{1}{2} \left[ K_1 (\partial_x \delta \hat{\mathbf N})^2 +  K_3 (\partial_z \delta \hat{\mathbf N})^2 \right] ,
\end{equation}
where $\hat{\mathbf N}$ denotes the layer normal, and $K_3 \propto \xi_D^2$. As discussed by Nelson \& Pelcovits~\cite{renorm}, non-linearities in the nematic free energy modify the nematic Frank constants $K_1$ and $K_3$ such that at scales longer than $d e^{\xi_D^2/a^2}$ the nematic can be described by a single Frank constant $\propto \xi_D^2$. In practice, this length scale can be very large compared to typical system sizes, so one usually sees a 2-Frank constant nematic. A study of the dynamics of smectic films, taking dislocations into account~\cite{Toner}, yields nematic behaviour corresponding to Eq.~(\ref{fn}) at long length scales, with a nematic kinetic coefficient that vanishes like $n_D \propto e^{-E_D/k_B T}$ at low temperatures.

\section{Smectic Hydrodynamics}
\label{normal}

The hydrodynamic variables for a two-dimensional smectic are the layer displacement $u$, and the conserved momentun densities $g_x$, $g_z$. In this section we focus for simplicity on the dynamics of free-standing smectic films~\cite{film}, where the momentum is conserved to a good approximation. The drag due to a liquid subphase is considered in Sec.~\ref{subphase}. We assume the smectic to be incompressible, and so neglect density fluctuations, setting the density $\rho = {\mathit const}$.

Consider the stress tensor $\sigma_{ij} = \nu_{ijkl} \partial_k g_l$. From the symmetry properties of the viscosity tensor, it can easily be argued that there should be 4 independent viscosity coefficients: $\eta_{xxxx}$, $\eta_{zzzz}$, $\eta_{xxzz} = \eta_{zzxx}$, and $\eta_{xzxz} = \eta_{zxzx} = \eta_{xzzx} = \eta_{zxxz}$. Upon denoting
\begin{equation}
\label{hu}
h \equiv - \frac{\delta {\mathcal F}}{\delta u} = B (\partial_z^2 - \lambda^2 \partial_x^4) u ,
\end{equation}
the equations of motion can be written as~\cite{prost}
\begin{mathletters}
\begin{eqnarray}
\label{dut}
\frac{\partial u}{\partial t} &=&  \frac{g_z}{\rho} + \lambda_p h \\
\frac{\partial g_z}{\partial t} &=& h - \partial_z p + \tilde{\nu_z} \partial_z^2 g_z + \nu \partial_x^2 g_z + \nu' \partial_z \partial_x g_x  \\
\frac{\partial g_x}{\partial t} &=& \mbox{\ } - \partial_x p + \tilde{\nu_x} \partial_x^2 g_x + \nu \partial_z^2 g_x + \nu' \partial_x \partial_z g_z
\end{eqnarray}
\end{mathletters}
where $\lambda_p$ is the permeation constant for the smectic, $p$ is the surface pressure, and we have switched to kinematic viscosities by dividing by the density, $\nu \equiv \eta/\rho$, and denoted the four viscosities $\nu_x, \nu_z, \nu$ and $\nu'$. The condition of constant density: $\frac{\partial \rho}{\partial t} = - \partial_x g_x - \partial_z g_z = 0$, can be used to reduce the number of viscosity coefficients to 3, and decouple the $g_x$-motion from $g_z$. Permeation refers to the dissipative mode of mass transfer in smectics where the molecules jump from layer to layer in order to relax a layer distortion.

Dislocations in the smectic introduce cuts into the displacement field, but it is possible to define the gradient ${\mathbf s} = {\mathbf \nabla} u$ as a single-valued quantity~\cite{Toner}. In the presence of dislocations, ${\mathrm s}_x$ and ${\mathrm s}_z$ are independent variables, with ${\mathbf \nabla} \times {\mathbf s} = - \hat{\mathbf y} d m({\mathbf r})$, where $m({\mathbf r})$ is the conserved dislocation density. Now
\begin{equation}
\frac{\partial {\mathbf s}}{\partial t} = {\mathbf \nabla} \frac{\partial u}{\partial t} + d \hat{\mathbf y} \times {\mathbf J}_D ,
\end{equation}
where ${\mathbf J}_D$, the dislocation current, which satisfies
\begin{equation}
\partial_t m + {\mathbf \nabla} \cdot {\mathbf J}_D = 0 ,
\end{equation}
is given by
\begin{equation}
{\mathbf J}_D = n_D \underline{\Gamma} \cdot {\mathbf f} - T \underline{\Gamma} \cdot {\mathbf \nabla} m .
\end{equation}
Here we have introduced the kinetic coefficients $\underline{\Gamma} = \left[\begin{array}{cc}\Gamma_x & 0 \\ 0 & \Gamma_z \end{array} \right]$, and the 2-d analog of the Peach-Koehler force ${\mathbf f} = d (B {\mathrm s}_z, B \lambda^2 \partial_x^2 {\mathrm s}_x)$; we have also set $k_B=1$ for convenience. We have imposed the Einstein relation which relates the mobility embodied in the first term to the diffusion constant implicit in the second through the common matrix $\underline{\Gamma}$. Since $\Gamma_z$ and $\Gamma_x$ correspond to dislocation glide and climb respectively, we expect $\Gamma_z \gg \Gamma_x$. The equations of motion are then:
\begin{mathletters}
\label{eqm}
\begin{eqnarray}
\frac{\partial g_z}{\partial t} &=& B (\partial_z {\mathrm s}_z - \lambda^2 \partial_x^3 {\mathrm s}_x) + \pi' + \nu \partial_x^2 g_z + \nu_z \partial_z^2 g_z \\
\frac{\partial {\mathrm s}_x}{\partial t} &=& \partial_x \frac{g_z}{\rho} + \lambda_p B (\partial_z^2 - \lambda^2 \partial_x^4) {\mathrm s}_x + \Gamma_z \left( n_D d^2 B \lambda^2 \partial_x^2 {\mathrm s}_x - T \partial_z (\partial_x {\mathrm s}_z - \partial_z {\mathrm s}_x) \right) \\
\frac{\partial {\mathrm s}_z}{\partial t} &=& \partial_z \frac{g_z}{\rho} + \lambda_p B (\partial_z^2 - \lambda^2 \partial_x^4) {\mathrm s}_z - \Gamma_x \left( n_D d^2 B {\mathrm s}_z - T \partial_x (\partial_x {\mathrm s}_z - \partial_z {\mathrm s}_x) \right)
\end{eqnarray}
\end{mathletters}
where we have used $\pi'$ to denote $-\partial_z p$, and $\nu_z = \tilde{\nu_z}-\nu'$.

If conservation of momentum is neglected~\cite{Toner}, Eqs.~(\ref{eqm}) lead in the limit of long wavelengths and low frequencies to a relaxation frequency for $s_z$ (which describes layer compression)
\begin{equation}
\omega_{s_z} = - i \Gamma_x n_D d^2 B ,
\end{equation}
and for ${\mathrm s}_x$ (\textit{i.e.}, layer undulations) a diffusive frequency
\begin{equation}
\omega_{s_x}({\mathbf q}) = - i \left((\Gamma_z n_D d^2 B \lambda^2) q_x^2 + (T \Gamma_z + \lambda_p B) q_z^2 \right) .
\end{equation}
Using the relation $\delta n = \partial_x u$~\cite{prost}, this last result corresponds to a nematic-like behaviour for the director $\hat {\mathbf N} = \hat {\mathbf z} + {\mathrm s}_x \hat {\mathbf x}$. Including $g_z$ in the hydrodynamic treatment introduces a pair of coupled $g$-$s_x$ modes with both diffusive and propagating characteristics. Denoting $\omega_{g_z}({\mathbf q}) = - i (\nu q_x^2 + \nu_z q_z^2)$, the coupled modes have characteristic frequencies
\begin{equation}
\omega({\mathbf q}) \approx \left(\frac{\omega_{g_z}+\omega_{s_x}}{2}\right) \pm \sqrt{\left(\frac{\omega_{g_z}-\omega_{s_x}}{2}\right)^2 + \frac{B}{\rho} \lambda^2 q_x^4} .
\end{equation}
Propagation dominates for ${\mathbf q} \parallel \hat {\mathbf x}$ if the dissipation is small enough, which is possible at low temperatures, leading to
\begin{equation}
\omega({\mathbf q}) = \pm \sqrt{\frac{B}{\rho}} \lambda q_x^2 .
\end{equation}

As in the case of hexatics, coupling to the smectic displacement field modifies the viscosity of the film. In the absence of dislocations, it is not possible to shear the smectic film without breaking it. The glide motion of dislocations facilitates shear deformation. Permeation can also support shear at short length scales. To calculate the effective viscosity, we apply an external shear stress to the system, and calculate the steady state response for $g_z$. The stress-strain relation calculated in Appendix \ref{eff} then yields
\begin{equation}
\label{eta}
\eta^{\mathit eff} = \eta \left( 1 + \frac{1}{\eta \Gamma_z n_D d^2} \right)
\end{equation}
and
\begin{equation}
\label{etaz}
\eta_z^{\mathit eff} = \eta_z \left( 1 + \frac{1}{\eta_z \Gamma_x n_D d^2} \right)
\end{equation}
in the long wavelength limit. This is similar in form to the viscosity correction for hexatics: $\eta \rightarrow \eta  \left( 1 + \frac{1}{4 \eta \Gamma_6}\right)$.

A similar calculation for  ${\mathrm s}_x$ and ${\mathrm s}_z$ (again using the method sketched in Appendix \ref{eff}) shows that
\begin{equation}
\Gamma_z n_D d^2 \rightarrow \Gamma_z n_D d^2 \left( 1 + \frac{1}{\eta \Gamma_z n_D d^2} \right)
\end{equation}
and
\begin{equation}
\Gamma_x n_D d^2 \rightarrow \Gamma_x n_D d^2 \left( 1 + \frac{1}{\eta_z \Gamma_x n_D d^2} \right)
\end{equation}
The permeation constant is not affected by the coupling.

At low temperatures (or large dislocation energy $E_D$), $n_D$ rapidly approaches 0 as $e^{-E_D/T}$, and the effective viscosity of the smectic begins to diverge as $e^{E_D/T}$, whereas the response of the smectic displacement strains ${\mathrm s}_x$ and ${\mathrm s}_z$ to an external force goes to the finite value $\eta^{-1}$ instead of vanishing as $\Gamma_z n_D d^2$. However, since the permeation mode relaxes shear over scales shorter than the permeation length $\delta = \sqrt{\lambda_p \eta}$, the divergence of the shear viscosity is cut off for $q_x \gg \sqrt{\eta \Gamma_z} (d/\delta \xi_D)$ according to
\begin{equation}
\Delta \eta \:q_x^2 = \frac{q_x^2}{\Gamma_z n_D d^2 + \lambda_p q_x^2} , \qquad{\mathrm or, }\qquad \eta \rightarrow \eta \left( 1 + \frac{1}{\eta \Gamma_z n_D d^2 + \delta^2 q_x^2} \right) .
\end{equation}
Since these hydrodynamic equations are valid only for wavelengths longer than the dislocation correlation length in the x-direction, i.e., $q_x \ll \xi_\perp^{-1}$ where $\xi_\perp = (\lambda \xi_D^2)^{1/3}$~\cite{Toner}, this rounding off of the viscosity will extend to the hydrodynamic range only if $\sqrt{\eta \Gamma_z} (d \lambda^{1/3} / \delta \xi_D^{1/3}) \ll 1$. We expect the bare viscosity $\eta$ and $\lambda = \sqrt{K_1/B}$ to stay finite as $T \rightarrow 0$. However, $\xi_D$ diverges as $e^{E_D/2T}$. We expect the permeation constant $\lambda_p$ to behave like $e^{-E_p/T}$ where $E_p$ is the energy barrier for molecules to jump from one layer to the next. The dislocation kinetic coefficient $\Gamma_z$ would similarly correspond to the activation energy $E_g$ for dislocation glide by breaking and reforming of bonds around the dislocation core. But this energy barrier should be small compared to that required for molecular hopping across the layers, and we shall ignore it in comparison. Then the above condition is satisfied provided $E_D/3 > E_p$, so that $\xi_D^{1/3} \rightarrow \infty$ faster than $\delta \rightarrow 0$.

A similar rounding off is possible in principle for $\eta_z$:
\begin{equation}
\eta_z \rightarrow \eta_z + \frac{1}{\Gamma_x n_D d^2 + \lambda_p q_z^2} ,
\end{equation}
however, this saturation is unobservable in the hydrodynamic limit because the dislocation correlation length diverges more strongly in the z-direction: $q_z^{-1} \gg \xi_\parallel = (\xi_D^4/\lambda)^{1/3}$. Since dislocation climb is similar to the permeation process, $\Gamma_x$ should behave like $\lambda_p$, and $q_z^2 \propto n_D^{4/3}$ would be much smaller than $n_D$ at low temperatures.

Although we have assumed the viscosity to be independent of shear rate, at high shear rates we must account for shear thinning brought about by the increase in unbound dislocations in the presence of the shear strain. The mechanism for dislocation proliferation under a shear stress is similar to that described by Bruinsma \textit{et al.}~\cite{shear-thinning} for a 2d crystal of point particles. The stress tilts the effective potential well binding the dislocation pair, allowing the pair to dissociate. The extra density of unbound dislocations facilitates further relaxation of the stress so that the effective viscosity decreases with increasing shear rate (the shear strain in the steady state depends on the shear rate imposed upon the flow). Note from Eqs.~(\ref{eta}) and (\ref{etaz}) that the effective viscosities do indeed drop with increasing dislocation density $n_D$.

The same mechanism would also apply to shear flow in a hexatic film where disclination unbinding would occur in the presence of a strain in the bond-orientation angle. Since the orientational order parameter is coupled to the flow as in Eqs. (\ref{hex}), disclinations can mediate the shear thinning mechanism in the hexatic phase.

\section{Channel flow of smectic films}
\label{channel}

We are interested in flow under shear or a pressure gradient for a film oriented with the layering direction along the channel (see Fig.~\ref{flow}). From the previous discussion, we expect a nematic-like profile for ${\mathrm s}_x$ unless the dislocation density is small, in which case we are in the permeation regime and shear is only supported in a boundary layer of width $\delta$. We assume the channel is much wider than the dislocation correlation length $\xi_D$, so that the hydrodynamic treatment is valid. Discussion of the effects of a subphase will be deferred to Sec.~\ref{subphase}.

In the steady state, we expect ${\mathrm s}_z$ to be constant, and the equations of motion (\ref{eqm}) reduce to
\begin{mathletters}
\begin{eqnarray}
\frac{\partial g_z}{\partial t} &=& \pi' + \nu \partial_x^2 g_z - B \lambda^2 \partial_x^3 {\mathrm s}_x \\
\frac{\partial {\mathrm s}_x}{\partial t} &=& \frac{g_z}{\rho} - \lambda_p B \lambda^2 \partial_x^4 {\mathrm s}_x + \Gamma_z n_D d^2 B \lambda^2 \partial_x^2 {\mathrm s}_x
\end{eqnarray}
\end{mathletters}
For convenience we reduce the variables by their dimensionless counterparts:
\begin{equation}
x \rightarrow a x, \quad g_z \rightarrow \left(\frac{\pi' a^2}{\nu}\right) g_z, \quad {\mathrm s}_x \rightarrow \left(\frac{\pi' a^3}{B \lambda^2}\right) {\mathrm s}_x ,
\end{equation}
where $2a$ is the channel width, and $x$, $g_z$ and $s_x$ are now dimensionless. We also define the dimensionless dislocation density
\begin{equation}
\Delta \equiv \eta \Gamma_z n_D d^2 ,
\end{equation}
and $b \equiv a/\delta$, the scaled channel width. In the presence of dislocations, it is convenient to define $b' \equiv b \sqrt{1 + \Delta}$.

Upon solving the equations above with the no-slip boundary condition: at $x = \pm a$, $g_z = 0$ and the permeation current $\propto h \propto \partial_x^3 {\mathrm s}_x = 0$ (see Eqs. (\ref{hu}) and (\ref{dut})), we find
\begin{equation}
g_z(x) = \frac{1}{1+\Delta} \left[ \Delta \frac{(1-x^2)}{2} + \frac{1}{b'^2} \left( 1 - \frac{\mbox{cosh}(b' x)}{\mbox{cosh}(b')} \right) \right]
\end{equation}
and
${\mathrm s}_x = \partial_x u$ where
\begin{equation}
u(x) = \frac{1}{1+\Delta} \left[ \frac{(1-x^2)^2}{24} + \frac{\mbox{tanh}b'}{b'^3} \frac{(1-x^2)}{2} + \frac{1}{b'^4} \left( 1 - \frac{\mbox{cosh}(b' x)}{\mbox{cosh}(b')} \right) \right] .
\end{equation}

\noindent
There are two regimes of interest here:
\begin{itemize}

\item \underline{narrow channel}: $b' \ll 1\quad$ ($\delta \gg a \sqrt{1+\Delta}$):
\[ g_z \rightarrow \frac{(1-x^2)}{2} , \]
i.e.\ we recover the usual Poiseuille profile expected for a structureless fluid. The permeation current $\lambda_p h$ has the same form, but is smaller than the momentum density by a factor of ${\mathcal O}(b^2)$. Also, in the same limit,
\[ u(x) \approx \frac{b^2}{720} (1-x^2)^2 (13-x^2) . \]
Since $u \ll 1$, the deviations in layer tilt, $\theta = -{\mathrm s}_x$, are small.

\item \underline{wide channel}: $b' \gg 1\quad$ ($a \sqrt{1+\Delta} \gg \delta$):

there are two distinct contributions to $g_z$ (Fig.~\ref{profile}):
\[ g_z \rightarrow \left( \frac{\Delta}{1+\Delta} \right) \left( \frac{1-x^2}{2} \right) + \frac{1 - e^{-b' (1-|x|)}}{b^2 (1+\Delta)^2} .\]
If $\Delta b'^2 \gg 1$ ($a \sqrt{\Delta (1+\Delta)} \gg \delta$), then the second term can be neglected and dislocations  restore a fluid like response, but with an effective viscosity
\begin{equation}
\label{etad}
\eta^{\mathit eff} = \eta \left(1+\frac{1}{\Delta}\right) ,
\end{equation}
confirming the result we found in the previous section. 
On the other hand, if the dislocation density is so small that $\Delta b'^2 \ll 1\quad$ ($a \sqrt{1+\Delta} \gg \delta \gg a \sqrt{\Delta (1+\Delta)}$), then the second term dominates and the plug flow profile characteristic of permeation flow can be seen.

In the wide channel limit, we also have
\[ u(x) \approx \frac{1}{\Delta} \frac{(1-x^2)^2}{24} + \frac{1}{(1+\Delta)^{5/2}} \frac{1}{b^3} \frac{(1-x^2)}{2} + \frac{1}{(1+\Delta)^3} \frac{1}{b^4} \left(1 - e^{-b' (1-|x|)}\right) .\]

\end{itemize}

\noindent
For the general case, we can estimate the effective viscosity from the flow rate: for Poiseuille flow, the momentum flux is given by $\int_{-a}^{a} g_z dx = \frac{2}{3} \frac{\pi' a^3 \rho}{\eta}$. Using this as the \textit{definition} of $\eta^{\mathit eff}$, we find
\begin{equation}
\frac{\eta}{\eta^{\mathit eff}} = \frac{\Delta}{1+\Delta} + \frac{3}{b'^2 (1+\Delta)} \left( 1-\frac{\mbox{tanh}b'}{b'} \right) .
\end{equation}
For $b' \gg 1$ (permeation regime),
\begin{equation}
\eta^{\mathit eff} = \eta \left(\frac{1+\Delta}{\Delta + 3/b'^2}\right)
\end{equation}
which reduces to (\ref{etad}) for $\Delta b'^2 \gg 1$, whereas for $\Delta b'^2 \ll 1$ (and hence $\Delta \ll 1$),
\begin{equation}
\frac{\eta^{\mathit eff}}{\eta} = \frac{b^2}{3} \qquad{\mathrm or}\qquad \eta^{\mathit eff} = \frac{a^2}{3 \lambda_p} ,
\end{equation}
reminiscent of the result $1/\lambda_p q_x^2$ we found for low dislocation densities in the previous section.

For $b' \ll 1$ ($b \ll 1$) (permeation regime), we have
\begin{equation}
\frac{\eta}{\eta^{\mathit eff}} = 1 + \frac{2}{5} b^2 \qquad{\mathrm or}\qquad \eta^{\mathit eff} = \eta + \frac{2}{5} \frac{a^2}{\lambda_p} ,
\end{equation}
which is a small correction due to the permeation boundary layer.

\section{Channel flow with a normal fluid subphase}
\label{subphase}

In practice~\cite{sub}, both the film and the barriers that restrict its flow to a channel geometry, float on a volume of fluid with viscosity $\eta_b$ and finite depth $H$ (Fig.~\ref{channel-sub}). If the dimensionless parameter~\cite{Stone}
\[ \Lambda = \eta/\eta_b H \]
is small, then, as for a normal film, the effect of this subphase can be neglected, and the analysis of the previous section is sufficient to describe the two-dimensional film flow. If the subphase drag cannot be neglected, the flow profile can be calculated through an analysis similar to Ref.~\cite{Stone}. We outline the steps here, and comment on the limiting cases. 

We consider a subphase extending from $y=0$ (surface with film) to $y=-H$ (bottom). Let $\vec{v}(x,y)$ be the ($z$-independent) velocity field describing the subphase, $\vec{v} \equiv (v_x,v_y,v)$. The velocity profile in the film itself is $v_0(x) \equiv v(x,y=0)$.

In steady state, the equation of motion for $\vec{v}$ in the bulk of the subphase is:
\[ (\partial_x^2 + \partial_y^2)\,\vec{\upsilon} = 0 . \]
The boundary conditions on the subphase are:
\begin{itemize}
\item $\vec{\upsilon} = 0\quad$ for $\quad x \rightarrow \pm\infty\quad$ or $\quad y=-H\quad$ or $\quad y=0, |x|>a$,
\item $v_x$ and $v_y$ $=0\quad$ for $\quad y=0, |x|<a\quad$ as well.
\end{itemize}
We assume that the subphase is incompressible ($\vec{\nabla} \cdot \vec{\upsilon} = 0$), which implies that $v_x$ and $v_y$ must be zero.

For the film, the equations of motion at the surface are modified by the subphase drag ($K_1 = B \lambda^2$):
\begin{mathletters}
\begin{eqnarray}
&\pi' + \eta \partial_x^2 v_0 - K_1 \partial_x^3 {\mathrm s}_x - \eta_b \partial_y v |_{y=0} = 0 ,& \\
&\partial_x (v_0 - \lambda_p K_1 \partial_x^3 {\mathrm s}_x) + \Gamma_z n_D d^2 K_1 \partial_x^2 {\mathrm s}_x = 0 .&
\end{eqnarray}
\end{mathletters}
Once again, we scale variables such that
\begin{equation}
x \rightarrow a x, \quad y \rightarrow a y, \quad v \rightarrow \left(\frac{\pi' a^2}{\eta}\right) v, \qquad{\mathrm and}\qquad \partial_x^3 {\mathrm s}_x \equiv \left(\frac{\pi'}{K_1}\right) v_p .
\end{equation}
as well as
\begin{equation}
\delta \rightarrow a \delta, \quad H \rightarrow a H .
\end{equation}
All quantities are now dimensionless. Since $v_p$ is proportional to the ``permeation current'', it obeys the same boundary conditions as $v_0$.

The equations of motion can now be written as
\begin{mathletters}
\begin{eqnarray}
\label{eqa}
& \partial_x^2 v + \partial_y^2 v = 0 \quad{\mathrm for}\quad -H<y<0 \quad{\mathrm with}\quad v = 0 \quad{\mathrm at}\quad y=0,-H & \\
\label{eqb}
& 1 + \left( \delta^2 \partial_x^2 - (1+\Delta) \right) v_p - \Lambda \partial_y v|_{y=0} = 0 & \\
\label{eqc}
& \partial_x^2 v_0 = (\delta^2 \partial_x^2 - \Delta) v_p &
\end{eqnarray}
\end{mathletters}
where $\delta$ and $\Delta$ were defined in Sec.~\ref{channel}.
Eq.~(\ref{eqa}) implies that $v$ must have the form
\begin{equation}
v(x,y) = \int_0^\infty\!\!\! dk\: \frac{A(k)}{\cos{k H}} \cos{k x} {\mathrm \;sinh}k(H+y) .
\end{equation}
In terms of the Fourier transform $A(k)$, the film velocity
\begin{equation}
v_0(x) = \int_0^\infty\!\!\! dk\: A(k) {\mathrm tanh}(k H) \cos{k x} .
\end{equation}
Using Eq.~(\ref{eqc}), we can express $v_p$ in terms of $v_0$:
\begin{equation}
v_p(x) = \int_0^\infty\!\!\! dk\: \Omega(k) cos{k x} \quad{\mathrm where}\quad \Omega(k) = A(k) {\mathrm tanh}(k H) \left(\frac{k^2}{\delta^2 k^2 + \Delta}\right) .
\end{equation}
Upon substituting these relations into Eq.~(\ref{eqb}), we obtain a relation for the Fourier transform $A(k)$:
\begin{equation}
\label{P}
1 = \int_0^\infty\!\!\! dk\: A(k) \left[ \left( \delta^2 k^2 + (1+\Delta) \right) \left(\frac{k^2}{\delta^2 k^2 + \Delta}\right) {\mathrm tanh}(k H) + \Lambda k \right] \cos{k x} .
\end{equation}
The boundary condition $v_0(x) = 0$ for $|x| > 1$ imposes another constraint on the $A(k)$:
\begin{equation}
\int_0^\infty\!\!\! dk\: A(k) {\mathrm tanh}(k H) \cos{k x} = 0 \quad{\mathrm for}\quad |x| > 1 .
\end{equation}
This can be satisfied by $A(k)$ of the form
\begin{equation}
A(k) {\mathrm tanh}(k H) = k^{1/2-\beta} \sum_{m=0}^\infty a_m J_{2m-1/2+\beta}(k)
\end{equation}
where $\beta$ can be chosen for convenience of computation.
If this form is substituted into Eq.~(\ref{P}), the x-dependence can be integrated out to yield an infinite set of linear equations for the coefficients $a_m$:
\begin{equation}
\sum_{m=0}^\infty a_m G_{mn}^\beta(\Delta, \delta, \Lambda, H) = \frac{\delta_{n 0}}{2^{\beta-1/2} \Gamma(\beta+1/2)} , \qquad n = 0,1,2,\ldots
\end{equation}
where
\begin{equation}
G_{mn}^\beta(\Delta, \delta, \Lambda, H) = \int_0^\infty\!\!\! dk\: G(k; \Delta, \delta, \Lambda, H) k^{1-2\beta} J_{2m-1/2+\beta}(k) J_{2n-1/2+\beta}(k) .
\end{equation}

The ``kernel'' for the smectic case,
\begin{equation}
\label{G}
G(k; \Delta, \delta, \Lambda, H) = k^2 \left( 1 + \frac{1}{\Delta + \delta^2 k^2} \right) + \frac{\Lambda k}{{\mathrm tanh}(k H)} ,
\end{equation}
differs from that for a structureless fluid by the term $\frac{k^2}{\Delta + \delta^2 k^2}$ (see Fig.~\ref{kernel}). This term reflects the correction to $\eta q_x^2$ we found in Section~\ref{normal}. As discussed there, the correction is small compared to the normal term $k^2$ for $\Delta \gg 1$, but can grow at low temperatures. If $\Delta \gg \delta^2 k^2$, this correction simply appears as an enhancement of the effective viscosity. However, when $\Delta \ll \delta^2 k^2$, the correction gives rise to a qualitative change in the velocity profile, characteristic of the permeation regime. The plug flow profile (Fig.~\ref{profile}) in this regime is similar to that seen in the case of a thin sublayer ($H \ll 1$), with most of the shear occurring in a boundary layer thickness $\delta$ (as opposed to $\sqrt{H/\Lambda}$ for the ``thin sublayer'' case).

When the subphase drag on the film is large ($\Lambda \gg 1$), the profile is very similar to that of a normal film, since the second term in the ``kernel'', which is independent of the film structure, dominates the flow. The profile is semi-elliptical when the subphase is deep ($H \gg 1$), and resembles plug flow for thin sublayers ($H \ll 1$).

\section{Summary}
\label{summary}

We have studied the hydrodynamics of two-dimensional smectics incorporating dislocations~\cite{Toner} in the context of shear flow across the layers. The behaviour resembles that of a nematic for length scales beyond the dislocation correlation length, with an effective viscosity that represents the role of dislocation motion in making shear possible. At smaller length scales, the permeation mode of smectics determines the shear response. These different regimes can be observed in channel flow under a pressure head where the channel width sets the observation length scale, provided the drag due to the subphase can be neglected. At small dislocation densities, the permeation mode determines the flow profile, which evolves from a parabolic profile for channels narrower than the permeation length to a plug-flow shape as the channel becomes much wider. In the latter case (the permeation regime), shear is supported only in a boundary layer of thickness equal to the permeation length, hence the effective viscosity as determined by the net flow rate across the channel grows as the square of the channel width. On the other hand, for large dislocation densities, the flow profile is again parabolic, but with the viscosity modified by the dislocation density.

The dislocation density in turn depends on the shear rate through the shear strain supported by the layers in steady state. Under this strain, dislocation pairs in the smectic unbind at a lower energy cost, increasing their equilibrium density and helping to further relax the imposed strain, resulting in a shear thinning effect. This effect has been calculated for a 2d crystal of point particles by Bruinsma \textit{et al.}~\cite{shear-thinning}, and would also be present in the hexatic phase, mediated by disclinations rather than dislocations. The flow results in steady state strains in the bond orientation order parameter, which are relaxed by disclination motion. It would be interesting to explore this mechanism for shear thinning for the smectic films discussed here. 

When the film flows on the surface of a fluid subphase, and drag from the subphase must be taken into account, the flow profile depends on the relative viscosities of the film and the subphase as well as on the channel width and the subphase depth. The analysis by Stone~\cite{Stone}, which is supported by experiments, predicts the evolution of the parabolic profile into a semi-elliptical or plug-flow profile, depending on whether the drag is due to the subphase viscosity or a shallow subphase. In the Introduction, we showed that the same results apply to a hexatic film if described by an effective viscosity incorporating the coupling to the bond-orientation order parameter. In the situations described above where the subphase drag dominates the flow, these results are also applicable to smectic fillms, since the modification to the ``flow kernel'' of a smectic film with respect to an isotropic film is decoupled from the terms describing the influence of the subphase. The subphase drag manifests itself at long length scales where the film structure is unimportant.

\acknowledgements

It is a pleasure to acknowledge helpful conversations with D. Schwartz. This research was supported by the National Science Foundation, through the MRSEC Program through Grant DMR-98-09363 and through Grant DMR-9714725.

\appendix

\section{Calculation of effective viscosity}
\label{eff}

The hydrodynamic equations of motion can be represented schematically as
\begin{equation}
- i \omega X({\mathbf q},\omega) = - {\mathbf L}({\mathbf q}) X({\mathbf q},\omega) + f^{\mathit ext}({\mathbf q},\omega)
\end{equation}
where $X \equiv (g_z,{\mathrm s}_x,{\mathrm s}_z)$, ${\mathbf L}$ is a hydrodynamic matrix (see Eq.~(\ref{L}) below), and $f^{\mathit ext} = (\partial_j \sigma_{zj}^{\mathit ext},0,0)$, $\sigma_{zj}^{\mathit ext}$ being the applied stress. Upon solving for the response to $\sigma_{zj}^{\mathit ext}$, we find
\begin{equation}
X({\mathbf q},\omega) = \left( - i \omega + {\mathbf L}({\mathbf q}) \right)^{-1} f^{\mathit ext}({\mathbf q},\omega) .
\end{equation}
In the limit $\omega \rightarrow 0$, this simplifies to $X({\mathbf q},\omega=0) = {{\mathbf L}({\mathbf q})}^{-1} f^{\mathit ext}({\mathbf q},\omega=0)$, or,
\begin{equation}
g_z({\mathbf q},\omega=0) = \left( {{\mathbf L}({\mathbf q})}^{-1} \right)_{g_z g_z} f^{\mathit ext}({\mathbf q},\omega=0) .
\end{equation}
Upon inverting this relation, we find
\begin{equation}
i q_j \sigma_{zj}^{\mathit ext}({\mathbf q},\omega=0) = \left( \left( {{\mathbf L}({\mathbf q})}^{-1} \right)_{g_z g_z} \right)^{-1} g_z({\mathbf q},\omega=0) .
\end{equation}
Upon comparing this result to the definition of the viscosity: $\sigma_{ij} \equiv \nu_{ijkl} \partial_k g_l$, we find the effective viscosity tensor $\nu^{\mathit eff}({\mathbf q})$.

\noindent
Upon writing
\begin{eqnarray}
\left( \left( {{\mathbf L}({\mathbf q})}^{-1} \right)_{g_z g_z} \right)^{-1}
&=& \frac{\mbox{Det} {\mathbf L}({\mathbf q})}{\mbox{Minor}_{g_z g_z} {\mathbf L}({\mathbf q})}
	\nonumber \\
&=& {\mathbf L}({\mathbf q})_{g_z g_z} + \frac{
{\mathbf L}({\mathbf q})_{g_z{\mathrm s}_x} \mbox{Minor}_{g_z{\mathrm s}_x}{\mathbf L}({\mathbf q})+
{\mathbf L}({\mathbf q})_{g_z{\mathrm s}_z} \mbox{Minor}_{g_z{\mathrm s}_z}{\mathbf L}({\mathbf q})}
	{\mbox{Minor}_{g_z g_z}{\mathbf L}({\mathbf q})} ,
\end{eqnarray}
we see that the first term simply yields the bare viscosity, and the second is the correction due to the coupling.

A similar calculation can be carried out for the effcetive response of the displacement gradients ${\mathrm s}_x$ and ${\mathrm s}_z$ to external forces.

The response matrix, ${\mathbf L}({\mathbf q})$, for a smectic is
\begin{equation}
\label{L}
\left[ \begin{array}{ccc}
(\nu q_x^2 + \nu_z q_z^2) & -i B \lambda^2 q_x^3 & -i B q_z \\
- i q_x /\rho & \lambda_p B (q_z^2 + \lambda^2 q_x^4) + \Gamma_z (n_D d^2 B \lambda^2 q_x^2 + T q_z^2) & - T \Gamma_z q_z q_x \\
- i q_z /\rho & - T \Gamma_x q_x q_z & \lambda_p B (q_z^2 + \lambda^2 q_x^4) + \Gamma_x (n_D d^2 B + T q_x^2)
\end{array} \right]
\end{equation}
We take $q_z = 0$ when calculating $\nu^{\mathit eff}$, and $q_x = 0$ in calculating $\nu_z^{\mathit eff}$, which leads to Eqs. (\ref{eta}) and (\ref{etaz}).



\begin{figure}
\caption[Channel flow geometry with a subphase]{Channel flow geometry with a subphase. The film and the barriers forming the channel are on the surface of a water table. A surface pressure gradient is applied to the film along the channel. On the right are the typical flow profiles at various depths.}
\label{channel-sub}
\end{figure}

\begin{figure}
\caption[Channel flow with surface pressure gradient]{Channel flow under the influence of a surface pressure gradient along the channel: the smectic layers are normal to the flow.}
\label{flow}
\end{figure}

\begin{figure}
\caption[Flow profiles in different limits]{In the wide channel/macroscopic limit, there are two contributions to flow: solid line represents the parabolic profile due to dislocation-assisted shear, whereas dotted line represents the plug-flow profile characteristic of flow in the permeation regime.}
\label{profile}
\end{figure}

\begin{figure}
\caption[Kernel for the channel flow geometry]{$G_{\mathit smectic-A}/G_{\mathit isotropic}$ as a function of the wave-number $k$ (solid line). $G(k)$ (see Eq.~(\ref{G})) is the ``kernel'' that determines the flow profile of the film on a fluid subphase in the channel flow geometry shown in Fig.~\protect{\ref{channel-sub}}. The channel width is scaled to 1. For large $k$, the structure of the film is important, and the ratio differs from 1 by the Lorentzian $1/(\Delta+\delta^2 k^2)$ (dotted line). For small $k$, the subphase drag (characterized by $\Lambda$ and $H$) dominates the flow, and the ratio approaches 1, leading to the hump at small $k$.}
\label{kernel}
\end{figure}



\newpage

\pagestyle{empty}
\setcounter{figure}{0}

\begin{figure}
\centering
\leavevmode
\epsfbox{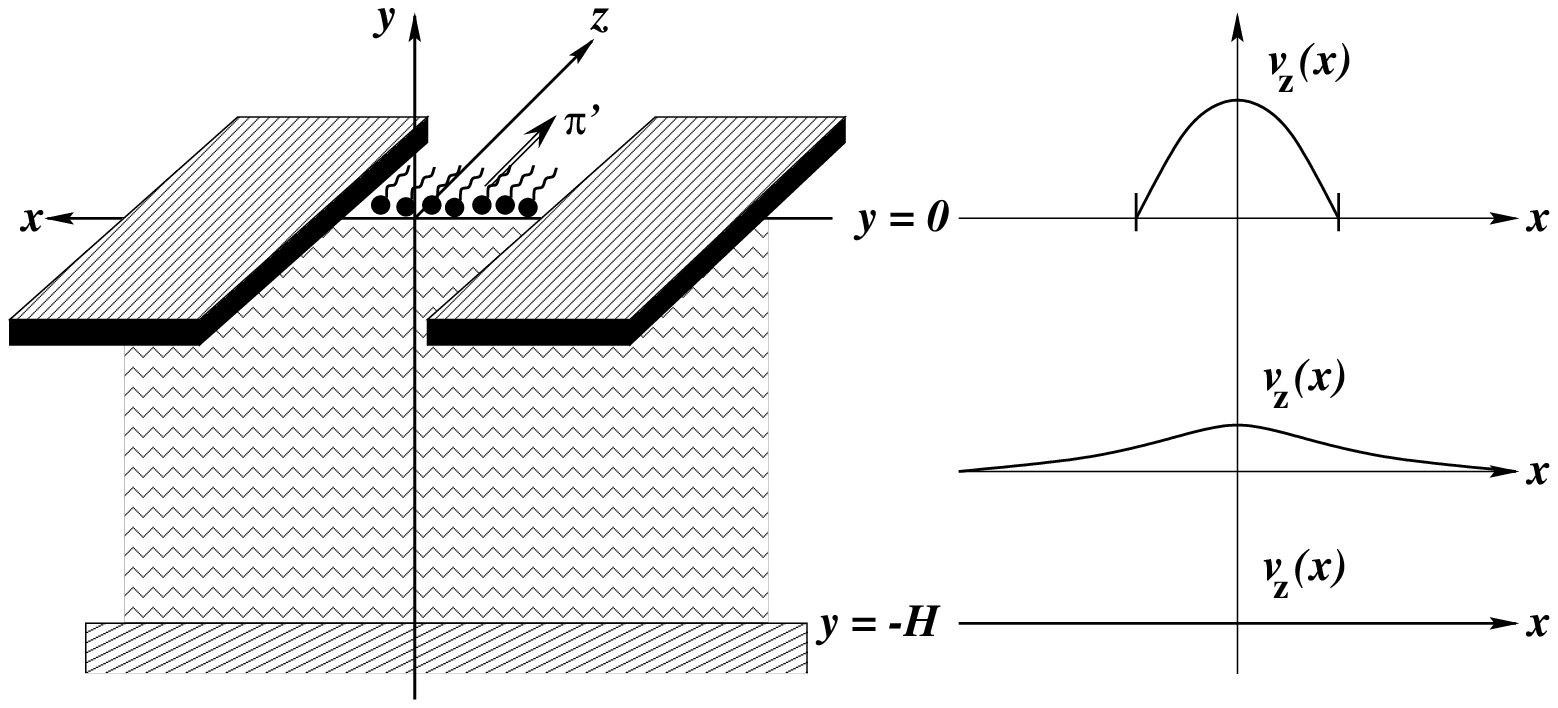}
\vfill
\caption{}
\end{figure}

\vspace*{\fill}

\begin{figure}
\centering
\leavevmode
\epsfbox{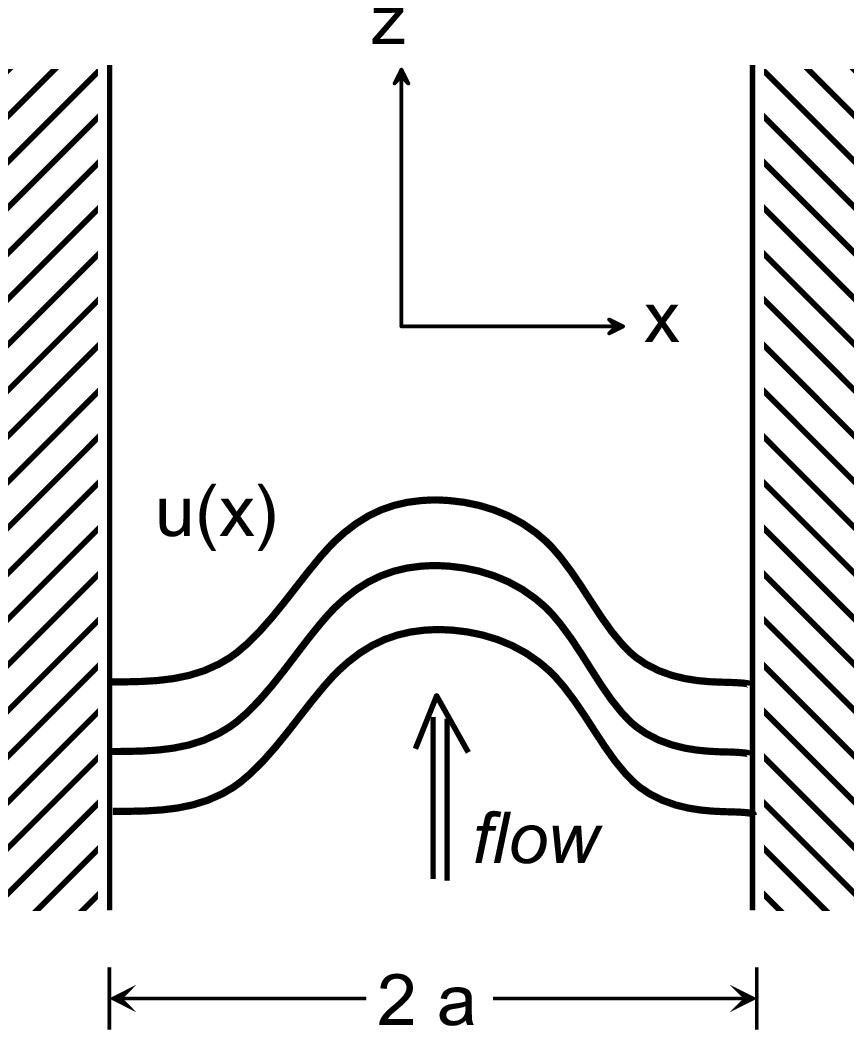}
\vfill
\caption{}
\end{figure}

\begin{figure}
\centering
\leavevmode
\epsfbox{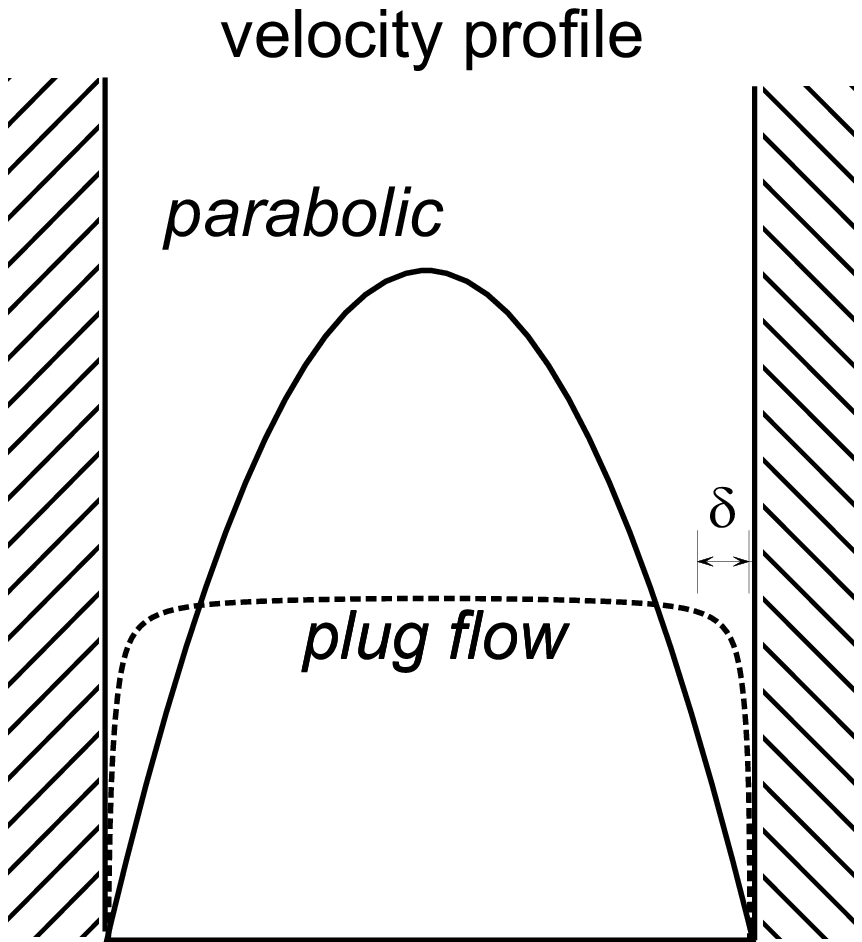}
\vfill
\caption{}
\end{figure}

\vspace*{\fill}

\begin{figure}
\centering
\leavevmode
\epsfbox{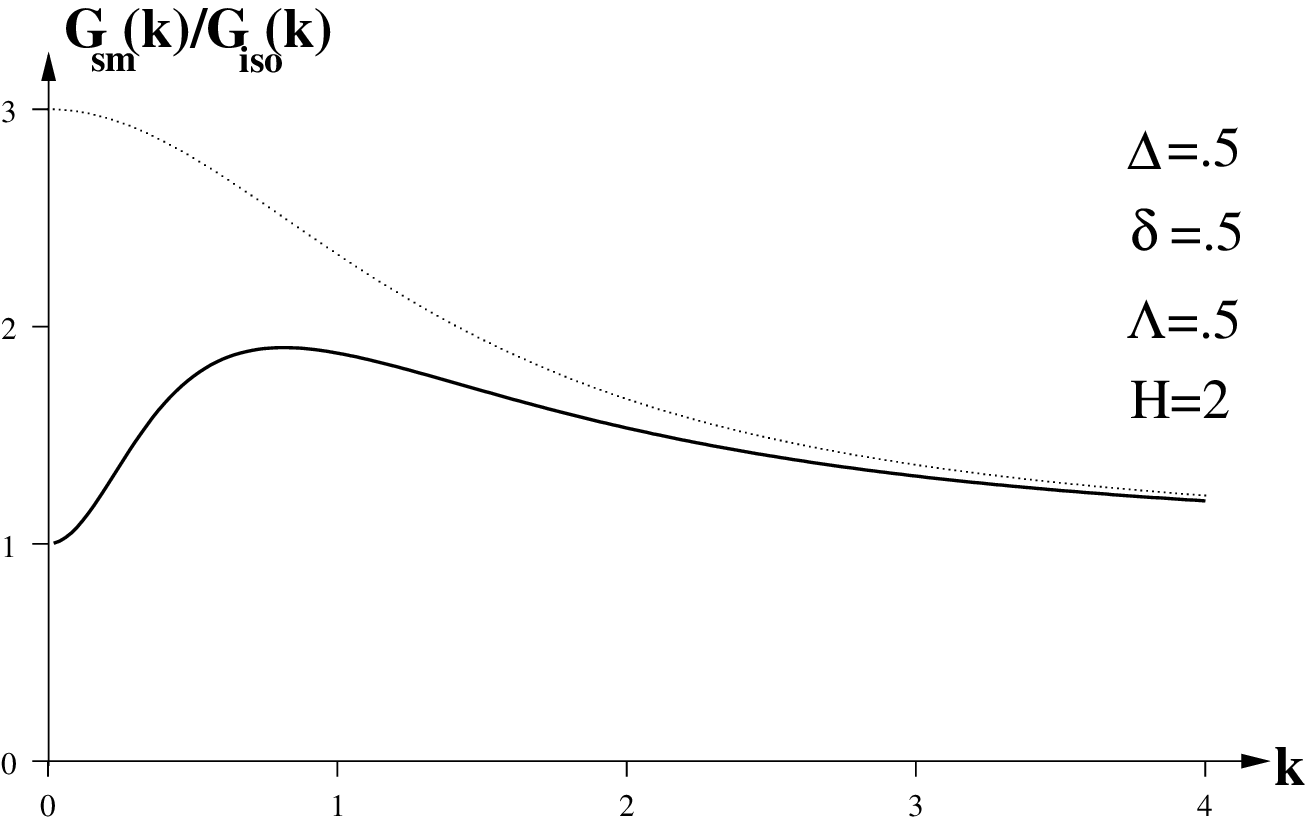}
\vfill
\caption{}
\end{figure}



\begin{thebibliography}{99}

\bibitem{orient} C. Mingotaud, B. Agricole, and C. Jego, J. Phys. Chem. \underline{99}, 17068 (1995).

\bibitem{Fuller} T. Maruyama, G. Fuller, C. Frank, and C. Robertson, Science \underline{274}, 233 (1996).

\bibitem{Schwartz} M. L. Kurnaz and D. K. Schwartz, Phys. Rev. E \underline{56}, No. 3, 3378 (1997).

\bibitem{elastic} R. S. Ghaskadvi, J. B. Ketterson, and P. Dutta, Langmuir \underline{13}, 5137 (1997).

\bibitem{shear-thinning} R. Bruinsma, B. I. Halperin, and A. Zippelius, Phys. Rev. B \underline{25}, No. 2, 579 (1982).

\bibitem{scresponse} For analogous phenomena in superfluid helium films, see V. Ambegaokar, B. I. Halperin, D. R. Nelson, and E. D. Siggia, Phys. Rev. Lett. \underline{40}, 783 (1978); V. Ambegaokar, B. I. Halperin, D. R. Nelson, and E. D. Siggia, Phys. Rev. B \underline{21}, No. 5, 1806 (1980).

\bibitem{private} D. K. Schwartz, private communication.

\bibitem{Toner} J. Toner and D. R. Nelson, Phys. Rev. B \underline{23}, No. 1, 316 (1981).

\bibitem{critical} F. J\"{a}hnig and F. Brochard, J. Physique \underline{35}, 301 (1974); F. Brochard, J. Physique Colloq. \underline{37}, C 3-85 (1976).

\bibitem{scaling} P. C. Hohenberg, B. I. Halperin, and D. R. Nelson, Phys. Rev. B \underline{22}, No. 5, 2373 (1980).

\bibitem{sub} D. K. Schwartz, C. M. Knobler, and R. Bruinsma, Phys. Rev. Lett. \underline{73}, 2841 (1994).

\bibitem{Stone} H. A. Stone, Phys. Fluids \underline{7}, 2931 (1995).

\bibitem{hydro} A. Zippelius, B. I. Halperin, and D. R. Nelson, Phys. Rev. B \underline{22}, No. 5, 2514 (1980).

\bibitem{prost} P. G. de Gennes and J. Prost, \textit{The Physics of Liquid Crystals} (Oxford Univ. Press, Oxford, 1993), 2nd ed.

\bibitem{gennes} P. G. de Gennes, Solid State Commun. \underline{10}, 753 (1972).

\bibitem{renorm} D. R. Nelson and R. A. Pelcovits, Phys. Rev. B \underline{16}, 2191 (1977).

\bibitem{film} H. T. Chiang, V. S. Chen-White, R. Pindak and M. Seul, J. Phys. II France \underline{5}, 835 (1995).

\end{thebibliography}
\end{document}